\documentclass[prd,superscriptaddress,twocolumn,amsmath,amssymb]{revtex4-1}
\usepackage{graphicx}
\usepackage{float}
\usepackage{color}
\usepackage{epstopdf,cancel,ulem}
\usepackage{epsf,latexsym,bbm,euscript}
\usepackage{amssymb,amsmath}
\usepackage{mathtools} 
\usepackage{times,graphics}
\usepackage{soul,xcolor}
\usepackage{epsfig}

\def\6{{\langle}}
\def\9{{\rangle}}

\newcommand{\be}{\begin{equation}}
\newcommand{\ee}{\end{equation}}
\newcommand{\ba}{\begin{eqnarray}}
\newcommand{\ea}{\end{eqnarray}}

\newcommand{\beq}{\begin{equation}}
\newcommand{\eeq}{\end{equation}}
\newcommand{\beqa}{\begin{eqnarray}}
\newcommand{\eeqa}{\end{eqnarray}}
\newcommand{\nn}{\nonumber}

\def\be{\begin{equation}}
\def\ee{\end{equation}}

\def\bali{\begin{align}}
\def\eni{{\end{align}}}

\def\nn{\nonumber}

\def\1{{{\mathbbm 1}}}

\def\half{\mbox{$1\over2$}}

\def\tit{{\tilde t}}

\def\pad{{\partial}}

\def\sg{\textsl{g}}

\def\Mc{M_\mathrm{c}}

\def\cO{\mathcal{O}}
\begin{document}

\title{ Role of evaporation in gravitational collapse}

\author{Valentina Baccetti}
\affiliation{Department of Physics \& Astronomy, Macquarie University, Sydney NSW 2109, Australia}
\author{Robert B. Mann}
\affiliation{Department of Physics and Astronomy, University of Waterloo, Waterloo, Ontario, Canada}
\affiliation{Perimeter Institute for Theoretical Physics, Waterloo, Ontario, Canada}
\author{Daniel R. Terno}
\affiliation{Department of Physics \& Astronomy, Macquarie University, Sydney NSW 2109, Australia}

\begin{abstract}

{We investigate the possibility that
 quantum effects responsible for black hole radiation do not allow for   horizon crossing of gravitationally collapsing matter
  in a finite time as seen by distant observers. We consider this in the context of the
collapse of evaporating massive thin dust shells  using two families of metrics to describe the exterior geometry: the outgoing Vaidya metric and the retarded  Schwarzschild metric. We describe how this hypothesis results in  a modified equation of motion for the shell. In each case  the collapse is accelerated due to evaporation, but the Schwarzschild radius  is not crossed.  Instead the shell is always at a certain sub-Planckian distance from this would-be horizon that depends only on the mass and evaporation rate, while a comoving observer encounters firewall-like energy density and flux with a natural cutoff.}
\end{abstract}
\maketitle

\section{Introduction} \label{intro}
  According to the clock of a distant observer (Bob at  spatial infinity),    collapse of a  matter distribution into a black hole takes an infinite amount of time. The final stage of a collapse is crossing of the event horizon --- boundary of the spacetime region from which no signal can escape. According to the comoving observer (Alice) it lasts only a finite amount of time. These are  the standard results of  classical general relativity \cite{ll2}.

{However the notion of an event horizon also plays a key role in quantum models of black holes \cite{kiefer:07,bd:82,modern,modern2}.}
  If Hawking radiation \cite{h:74,bd:82,modern,modern2} {exists} \cite{heffler:03} and its  late-time effects are described by  Page's formula \cite{page:76,modern,modern2}, only a finite amount of Bob's time elapses until the black hole  evaporates.
Black hole radiation can be derived in a number of ways \cite{bd:82,modern,modern2,ab:05}, {each  typically based} on three basic assumptions \cite{ab:05}: (i) the gravitational field is treated as a classical background; (ii)  the backreaction of the  created matter on the spacetime geometry is neglected, at least at the first stage of the analysis; (iii) field(s) of the emitted radiation are distinct from the collapsing matter. Positivity of the resulting steady-state energy flux at infinity, conservation of energy, and the relationships between event horizon area and mass for stationary black holes lead to a conclusion of a steady decrease of the horizon and provide a way to describe the changing spacetime geometry.

Hawking radiation precipitates the black hole information problem. Crossing of the event horizon by the collapsing matter and restoration (or the alleged non-restoration) of correlations during the evaporation provide   ingredients for the paradox \cite{kiefer:07, modern,modern2,mathur:09,ab:05,wald:01lrr,visser:08,new}. We focus on consistency of these two elements  since there is    an apparent causal contradiction: infinite collapse time vs. finite  subsequent evaporation time as perceived by Bob.

There are two alternatives for the final stages of the collapse: either  quantum effects
 facilitate a finite-time collapse --- crossing of a suitably defined {event} horizon within a finite amount of Bob's time --- or they {do} not. Both can restore the causal order.

If quantum effects lead to a finite-time collapse (according to the distant Bob), then there is no logical necessity in emitting  pre-Hawking radiation before the event horizon is formed.   Nonetheless, a spacetime region containing the horizon of  even a large  black hole would  still  be   non-classical. The collapse speed-up implies that either  a test particle that trails just behind the collapsing matter will also cross the fully formed horizon in a finite {coordinate} time,  or the ``original"  matter for some reason behaves differently from everything else.  Since derivations of the Hawking radiation do not assume exotic background structures \cite{h:74,bd:82,modern,modern2},  investigations of its logical consequences \cite{mathur:09} should not assume them either.

If  quantum effects   do not facilitate   {such} collapse, evaporation must start when the collapsing matter is sufficiently close to its Schwarzschild radius. Arguments that the collapse according to Bob is never complete and overlaps with the onset of Hawking radiation that begins when the collapsing matter concentrates near the gravitational radius, have been made before \cite{visser:08,ger:76,haj:86}. Numerical studies of   collapsing shells \cite{acvv:01} and  analytic  results \cite{liberati,vsk:07,kmy:13,ho:16a}  support this idea.

Accepting  the overlap between   collapse and evaporation presents the same alternative:  either quantum effects responsible for black hole radiation allow for the horizon crossing
(however properly defined in the quantum case) in finite time
for Bob or they do not.  In the latter case  the alleged loss of information, at least in a sense of its disappearance through a horizon \cite{h:76,wald:01lrr,modern,modern2},  never arises \cite{ho:16a,thes,pt:04,d:10}. 

We explore the consequences of {   pre-Hawking radiation} and present several  consistent scenarios where  the second possibility --- no horizon --- is realized. In Section II we discuss    considerations leading to our models, their specific assumptions as well as review the classical aspects of a thin shell collapse. Section III presents several detailed examples.  We conclude with the discussion of their applicability, generality and implications in Section IV. Appendices provide all the necessary calculational details.
We use $(-+++)$ signature of the metric and set $c=\hbar=G=k_B=1$.


\section{General considerations}
There is a useful
hierarchy of models that describe the interplay between
quantum mechanics and gravity. Relevant models begin with  quantum field  theory   on a curved
background and  advance through   semiclassical stochastic gravity \cite{sto-g} and
different effective field theories of matter-gravity
systems \cite{eft-g} to a full theory of quantum gravity in whatever form it takes \cite{kiefer:07}. Black holes   are discussed at each level of this hierarchy.

 Appearances of strongly non-classical regions, both deep within appropriately defined quantum black holes  and outside the Schwarzschild radius \cite{ab:05,amps,hr:15} are expected. Nevertheless, the classical event horizon  and quantum states that are associated with it play an important role in the quantum black hole models. Moreover,   whatever reservations can be raised against the semiclassical picture of black hole creation, radiation and evaporation, they are still the only ones that are fully solvable.  Hence investigations of the event horizon crossing within the semi-classical framework are important to establish its consistency and delineate its limitations.

Our treatment of this model is based on the following assumptions:
\begin{enumerate}
\item The classical spacetime structure is still meaningful and is described by a metric $g_{\mu\nu}$.
\item Classical concepts, such as trajectory, event horizon or singularity can be used.
\end{enumerate}
Assumption 1 leaves out the spacetime fluctuations. Within this framework there are no restrictions on the applicability of   classical concepts even, e.g., within the sub-Planckian distance from the Schwarzschild radius. While this is a stronger assumption than just   validity of the classical description outside the stretched horizon \cite{modern,modern2, amps}, it is  consistent with the goal of testing   accessibility of an infinitely sharp classical event horizon. Moreover, while this assumption is not usually emphasized, it underlines many discussions of the Hawking radiation and related problems. Specifically, the standard Penrose diagram of the creation of a black hole from the infalling matter with subsequent evaporation makes sense only if both Assumptions 1 and 2 are accepted \cite{bht:17}.

{We furthermore assume}
\begin{enumerate}
\setcounter{enumi}{2}
\item The collapse leads to a pre-Hawking radiation.
\item The metric is modified by quantum effects. The resulting curvature satisfies the semiclassical~equation
\be
R_{\mu\nu}-\half R g_{\mu\nu}=8\pi \6\hat{T}_{\mu\nu}\9, \label{semiein}
\ee
where $R_{\mu\nu}$ is the Ricci tensor corresponding to the metric $g_{\mu\nu}$ and $\6\hat{T}_{\mu\nu}\9$ is the expectation value of the stress-energy tensor.

\end{enumerate}

Assumption 3 is the direct consequence of our discussion in Section~\ref{intro}. In principle we make no assumptions about the specific form of the pre-Hawking radiation. It should be obtained as a part of a self-consistent analysis.
The expectation value of the stress-energy tensor should include all matter fields as well as gravitons, thus relaxing assumptions (ii) and (iii) from Section~\ref{intro}. Formally it is given by
\be
\6\hat{T}_{\mu\nu}\9=\frac{2}{\sqrt{-g}}\frac{\delta W}{\delta g^{\mu\nu}},
\ee
where $g=\det g_{\mu\nu}$ and $W$ is the effective action of quantum fields \cite{bd:82}, but only partial results are known for dynamical spacetimes.

Below we illustrate the    consequences of the Assumptions 1-4 in two examples. We discuss a generic spherically-symmetric setting in \cite{us-new}.

\subsection{The model}
 We consider  collapse  of a  massive  spherically-symmetric thin {dust} shell $\Sigma$
 \cite{poisson} in $D+1$ dimensional spacetime, $D\geq3$. Appendix~\ref{appa} summarizes the relevant conventions and definitions.

The spacetime inside the shell is flat \cite{poisson}. In  the absence of Hawking radiation  the metric
 \begin{eqnarray}
 ds^2_+ &=&- f(r_+) dt^2_+ + f(r_+)^{-1}dr^2_++r^{2}_+ d\Omega_{D-1}  \label{met0}\\
 &=& -f(r_+) du_+^2-2du_+ dr_+ + r_+^{2} {d\Omega_{D-1}}, \label{met01}
  \label{met2}
 \end{eqnarray}
where $f(r)=1-C/r^{D-2}$ {describes the exterior geometry in terms of standard coordinates $(t_+,r_+)$ and  outgoing Eddington-Finkelstein coordinates $(u_+,r_+)$. Here and in the following $dt _+ = du_+ +dr_+/f(r_+)$, and $d\Omega_{D-1}$ is the spherical volume element. Their counterparts in the Minkowski spacetime  inside the shell are $(t_-,r_-)$ and $(u_-,r_-)$, where $u_-=t_- - r_-$.
  The Schwarzschild radius   $r_\sg:=C^{1/(D-2)}$ \cite{hor:12} is the solution of $f(r)=0$. In three spatial dimensions $C=2M=r_\sg$.
}

The shell's trajectory is parameterized by its proper time $\tau$ (time of a comoving observer Alice) as $\big(T_\pm(\tau), R_\pm(\tau)\big)$ or $\big(U_\pm(\tau), R_\pm(\tau)\big)$ using, respectively, $(t,r)$ or $(u,r)$ coordinates outside and inside the shell.

The first junction condition \cite{poisson, isr:66}, which is the statement that the induced metric $h_{ab}$ is the same on the both sides of the shell $\Sigma$, $ds^2_\Sigma=h_{ab}dy^ady^b=-d\tau^2+R^2d\Omega_{D-1}$, leads to the identification $R_+\equiv R_-=:R(\tau)$. Trajectories of shell's particles are timelike, so
\be
\dot T_+=\sqrt{F+\dot R^2}/F,
\label{tdot}
\ee
{where the dot denotes a $\tau$ derivative and}
\be
\dot U_+=\frac{-\dot R+\sqrt{F+\dot R^2}}{F},
\label{udot}
\ee
where $F=1-C/R^{D-2}$.
The surface stress-energy tensor for a thin  dust shell   is
\be
S^{ab}=\sigma v^a v^b=\sigma \delta^a_\tau\delta^b_\tau,
\ee
where $\sigma$ is the surface density and $v^a$ are the components of the proper velocity  in the surface coordinates $y^a$. Discontinuity of the extrinsic curvature $K_{ab}$ is described by the second junction condition \cite{poisson,isr:66}
\be
S_{ab}=-\big([K_{ab}]-[K]h_{ab}\big)/8\pi, \label{eqofmot}
\ee
where {$K:=K^a_{\,a}$, and }$[K]:=K|_{\Sigma^+}-K|_{\Sigma^-}$ is the discontinuity of the extrinsic curvature $K$ across the two sides $\Sigma^\pm$ of the surface.
The equation of motion for the   shell can be written as
\begin{eqnarray}
  {\mathcal{D}(R) } &:=&  \frac{2\ddot R + F'}{2\sqrt{F+\dot R^2}} - \frac{\ddot R}{\sqrt{1+\dot R^2}} \nonumber\\
&& + (D-2)\frac{\sqrt{F+ \dot R^2} - \sqrt{1+\dot R^2}}{R} = 0 \label{sangC}
\end{eqnarray}
where  $\dot A=dA/d\tau$, $A'=\pad A/\pad r |_\Sigma$. This equation is  simple enough to have an analytic solution $\tau(R)$, {leading to the finite crossing time $\tau(r_\sg)$.} Appendix~\ref{curvature-v} presents details of the derivation of the equations of motion.

 Unlike previous investigations of  thin shell collapse \cite{dfu:76,h-shell,ss:15}, we are not dealing here with its influence  on a quantum field, but rather focus on the effects of  the resulting radiation on the shell dynamics. A metric  describing the geometry outside the evaporating shell is self-consistently determined from the Einstein equations (Assumption 4).  Using a null shell \cite{kmy:13,h-shell,hr:15} simplifies the analysis.  Here we investigate a massive shell, with the advantage of being able to   consider its evolution in the proper reference frame of Alice in addition to the asymptotic frame of Bob.  The proper reference frame  is crucial for establishing the physical meaning of the horizon.

We illustrate the general approach that was outlined in this section by two particular scenarios. The first scenario models the geometry outside the shell using the outgoing Vaidya metric \cite{kmy:13,vai:51,fw:99,ft:08}, which is a popular model of exterior geometry of evaporating black holes.   Apart from two obvious constrains we do not specify the mass function $C(u)$.  Next we introduce the retarded Schwarzschild metric, which  allows to describe evaporation from the point of view of Bob.

By monitoring
\be
x:=R-r_\sg,
\ee
 where $R(\tau)$ is determined by the equations of motion based on the assumed outside metric, we analyse  how evaporation modifies the classical shell dynamics.


\section{Examples}

\subsection{Vaidya metric}

The outgoing Vaidya metric   is  {given by \eqref{met2}
but with  $f\to f(u,r)=1-C(u)/r^{D-2}$, where we only assume that $f(u,r)>0$ for $r>r_\sg(u)\geq 0$ and $dC/du\leq0$}.
Here and in the following we drop the subscript ``+" from the variables.

The equation of motion of the shell  {\eqref{sangC} becomes} (see Appendix~\ref{curvature-v}),
\begin{eqnarray}
{\mathcal{D}(R)} -F_U \dot{ U} \left( \frac{\dot R}{2F\sqrt{F+\dot R^2}} -\frac{1}{2F}\right)= 0,
\label{eq:eom-vaidya}
\end{eqnarray}
where
$A_u:=dA/du$, and $\dot U$ is given by Eq.~\eqref{udot}. Solving Eq.~\eqref{eq:eom-vaidya} for $\ddot R$, then expanding in   inverse powers of $x$ and $C$
we obtain
\be
\ddot R = \frac{2 \dot R^2 \sqrt{1 + \dot R^2} \,  }{(D-2)^2 C^{\frac{D-4}{D-2}} (\dot R + \sqrt{1 + \dot R^2}) \, x^2}\frac{d C}{d U} + \cO(x^{-1}), \label{rddap}
\ee
where close to the Schwarzschild radius
\be
F\approx \frac{(D-2) x}{C^{1/(D-2)}} \equiv \frac{(D-2) x}{r_\sg}. \label{fclose}
\ee
{The coefficient of $dC/dU$ in \eqref{rddap} is positive, so} $\ddot R<0$. Hence sufficiently close to the Schwarzschild radius the collapse is always accelerated. At  later stages (when $|\dot R|\gtrsim2$) it can be approximated as
\be
\ddot R \approx \frac{4 \dot R^4}{(D-2)^2 C^{\frac{D-4}{D-2}}  \, x^2}\frac{d C}{d U} < 0.
\ee

We now evaluate {the shell's rate} of approach to $r_\sg$. {In its vicinity Eq.~\eqref{udot} reduces to $\dot U \approx-{2 \dot R}/{F}$. Using this expression and Eq.~\eqref{fclose} in the chain rule evaluation of $\dot r_\sg$ we find that
\be
\dot x \approx \dot R \left(1-\frac{2r_g}{(D-2)x}\frac{dr_\sg}{dC}\left|\frac{dC}{dU}\right|\right).
\ee
Hence a natural time-dependent scale for this problem is
\be
\epsilon_*:=\frac{2}{(D-2)^2C^{(D-4)/(D-2)}} \left|\frac{d C}{d U}\right|,
\ee
and
\be
\dot x\approx \dot R\big(1-\epsilon_*(\tau)/x(\tau)\big). \label{xeps}
\ee
Then the gap decreases only as long as $\epsilon_*<x$. If this is true for the entire duration of evaporation, we have $R>r_\sg$ until the evaporation is complete. Otherwise, once the distance between the shell and the Schwarzschild radius is reduced to $\epsilon_*$, it cannot decrease any further.}

From the point of view of Alice the collapse accelerates, while for Bob the shell is still stuck within a slowly changing coordinate distance $\epsilon_*$ from the slowly receding Schwarzschild radius. {Furthermore using \eqref{udot} it is straightforward to show that
close to $r_\sg$
\be
\dot C\approx 2C\frac{dC}{dU}\frac{|\dot R|}{x}
\ee
and so Alice will see  the evaporation rate $\dot C$ vanish if and only if Bob does; if  $dC/du\rightarrow0$ then~\eqref{eq:eom-vaidya} reduces to Eq.~\eqref{sangC},  $\epsilon_*\rightarrow 0$, and an horizon forms. If $dC/du< 0$ then since the gap $x$ {never vanishes,  $\dot R$ increases} but remains finite for finite $x$.

The only non-vanishing component of the stress-energy tensor is
\be
8\pi T_{uu}=-\frac{1}{r^2}\frac{dC}{du}.
\ee
 The energy density in the frame of an observer moving with a four-velocity $v^\mu$ is
\be
\rho=T_{\mu\nu}v^\mu v^\nu, \label{dens}
\ee
hence on the outer shell surface where
\be
\dot U\approx -2 \dot R/F\approx-2 \dot R C/x \label{uappro}
\ee
 we have
\be
\rho\approx \frac{1}{2\pi}\left|\frac{d C}{dU}\right|\frac{\dot R^2}{x^2}.
\ee
The energy flux in this frame is given by
\be
j_n:=T_{\mu\nu}v^\mu n^\nu=\rho, \label{flux}
\ee
where $n^\nu$ is given by Eq.~\eqref{outnorm}.see that the observer
Using Eq.~\eqref{rddap} it is possible to see that Alice will experience an enormously increased energy density and flux.  Approximating $x\sim\epsilon_*=2C|dC/dU|$ the asymptotic form of the equation of motion becomes
\be
\ddot R\approx\frac{\dot R^4}{C C_U}.
\ee
Hence  Alice will see the radiation flux   growing to some maximal value
contingent on the properties of the shell.   This blue-shift behaviour is a known feature of the standard black hole models \cite{pz:84} and provides a possible realization of a firewall acting via ``internal conversion" \cite{z-ap}. It could be regarded as a firewall with a natural upper cut-off.


\subsection{Retarded Schwarzschild metric}

 We introduce an alternative modification of the metric \eqref{met0} that describes spherically-symmetric evaporation. This example demonstrates that horizon avoidance is not limited to the Vaidya metric.
 We incorporate an   mass function $C(t)$  (as described by Bob), $C_B(t)$, into a local causality-respecting metric   via the retarded time, similarly to the derivation of the  Lienard-Wiechert potential in the radiation problem \cite{ll2}. As a result we can illustrate the analysis by numerical simulation of the process in Section~\ref{numerics}.

Outside the shell (the region  $(t,r):~r>R(t)$) we assume the  metric
\be
ds^2_+=-\tilde f(t,r) dt^2+\tilde f(t,r)^{-1}dr^2+r^2 d\Omega_{D-1}. \label{met1}
\ee
 This is the minimal modification of Eq.~\eqref{met0} consistent with the assumption   that an outgoing  {massless  particle propagates} in the Schwarzschild spacetime that is 
 frozen at the moment of its emission $\tit$, i.e.
\be
\frac{dr_\mathrm{out}}{dt}={1- {C(\tit\,)}/{r_\mathrm{out}^{D-2}}}.
\ee
Here $\tilde f(t,r)=1-C(\tilde t)/r^{D-2}\equiv f(\tit,r)$, and   the retarded time $\tit(t,r)$   is given by the implicit equation
\be
t-\tit=\int_{R(\tit)}^r\!\frac{dr'}{f(\tit,{ {r^\prime}})}\equiv r_*-R_*(\tit), \label{retarded}
\ee
where the tortoise coordinates are defined using the mass parameter at $\tit$. Outgoing null geodesics are the lines of $\tit=\mathrm{const}$.   We introduce $C(\tit)$ via  the implicit relationship \eqref{retarded}, $ {C(\tit):=C_B(t-r_*+R_*)}$ ,
 which in particular implies
\be
\frac{dC}{d\tit}=\left.\frac{dC_B(z)}{dz}\right|_{z=\tit}.
\ee
It is straightforward to show (Appendix \ref{ineq}) that for any choice of the function $C$ this metric is inequivalent to the outgoing Vaidya metric.

In this model the shell follows the classical trajectory $(T(\tau), R(\tau))$ until  a certain coordinate distance $\epsilon$  from the Schwarzschild radius, when the evaporation abruptly starts.  While definitely not an instantaneous event, the ignition time can be reasonably well-defined in, e.g.,  the adiabatic approximation \cite{bbg:11}, and we shall  set $\tau=t=0$ at this event for convenience.
Writing $F:=\tilde f(T,R)$,
the equation of motion \eqref{sangC}
for the collapsing and evaporating shell  now becomes (Appendix \ref{rets})
\be
  {\mathcal{D}(R) } - \frac{F_T \dot R}{F^2}= 0
\label{sang00}
\ee
where
 \begin{align}
& F'=\frac{C(D-2)}{R^{D-1}}-\frac{dC}{d\tit}\left.\frac{\pad \tit}{\pad r}\right|_\Sigma\frac{1}{R^{D-2}},\\
& F_T=-\frac{dC}{d\tit}\left.\frac{\pad \tit}{\pad t}\right|_\Sigma \frac{1}{R^{D-2}}.
\end{align}
The derivatives of the retarded   time on the shell are
\be
\left.\frac{\pad \tit}{\pad t}\right|_\Sigma=\frac{1}{1-R_{T}/F}, \qquad \left.\frac{\pad \tit}{\pad r}\right|_\Sigma=-\frac{1}{F-R_{T}},
\ee

 Eq.~\eqref{fclose} holds apart from the last stages of evaporation (where $x(\tau)\sim C(\tau)$). Hence \eqref{tdot} becomes
\be
{\dot T\approx +\frac{r_\sg}{(D-2)x}|\dot{R}|}.
\ee
Noting that
\be
\dot R=\frac{dR}{dt}\dot T\approx -\frac{dR}{dt}\frac{\dot R}{F},
 \ee
 implies the equality $R_\tit|_\Sigma=R_t\equiv R_T\approx-F$, the evaporation parameterized by the shell's proper time is given by
\begin{eqnarray}
\frac{dC}{d\tau}&=&\frac{dC}{d\tit}\left(\left.\frac{\pad \tit}{\pad t}\right|_\Sigma\dot T+\left.\frac{\pad \tit}{\pad r}\right|_\Sigma\dot R\right) \label{cexact} \\
&\approx&-\frac{dC}{d\tit}\frac{\dot R}{F}=-\left.\frac{dC(T)}{dT}\right|_{T=T(\tau)}\frac{\dot R}{F}
 \label{evaptau}
\end{eqnarray}
since at short distances from the Schwarzschild radius ${\pad \tit}/{\pad t}|_\Sigma\approx\frac{1}{2}$ and $  {\pad \tit}/{\pad r} |_\Sigma\approx- {1}/{(2F)}$.

Solving  Eq.~\eqref{sang00} for $\ddot R$ and expanding in inverse powers of $x$ and $C$  we obtain
\be \label{ddotRa}
\ddot R = \frac{\dot R^2\sqrt{1+\dot R^2}}{2D(D-2)^2C^{\frac{D-4}{D-2}}\big(\dot R+\sqrt{1+\dot R^2}\big)x^2}\frac{dC}{dT} +\cO(x^{-1}),
\ee
for the dominant term  of the shell's acceleration as it approaches the Schwarzschild radius, {and the expression further simplifies (Eq.~\eqref{domi})} for $|\dot R|\gtrsim2$.

Approach to the shell $\dot x=\dot R-\dot r_\sg$ is still governed by Eq.~\eqref{xeps}.
 Using Eq.~\eqref{evaptau} we find that  now the  distance scale is
\be
\epsilon_*(\tau)=\frac{1}{(D-2)^2C^{(D-4)/(D-2)}} \left|\frac{d C}{d T}\right|. \label{epstars}
\ee

As an illustration we consider the  late-time evaporation law
 \be\label{PageF}
\frac{dC_B}{dt}= {-}\frac{D-2}{D\kappa}\frac{1}{ C_B^{2/(D-2)}},
\ee
 where the constants are discussed in Appendix \ref{page-d}. Starting from the initial value $C_0$ the evaporation lasts a finite time $t_E$,
\be
C_B(t)=C_0\left(1- {t}/{t_E}\right)^{\frac{D-2}{D}},  \qquad t_E=\kappa C_0^{\frac{D}{D-2}}. \label{evaplaw}
\ee

Using \eqref{PageF} the asymptotic dynamics of the evaporating  shell  is given by the system
\begin{align}
&\ddot R\approx -\frac{\dot R^4}{D(D-2)\kappa Cx^2},  \label{domi} \\
&\dot C\approx\frac{1}{D\kappa C^{1/(D-2)}}\frac{\dot R}{x},  \label{domi2}
\end{align}
where \eqref{domi} is valid for large $|\dot R|\gtrsim 2$.

We see that the effect of evaporation is negligible until  about $\tau_1\sim\epsilon/|\dot R(0)|$, when the distance to the Schwarzschild radius reaches $x=\cO(\epsilon_*)$.  Once it is reached
we are guaranteed that $\dot x>0$ and so the shell does not cross the Schwarzschild radius  {despite
rapidly increasing acceleration.} While our choice of the switching-on parameters determines the value of $\tau_1$, the steady-state coordinate distance $\epsilon_*$ depends only on the system properties.


\subsection{Numerical solutions}\label{numerics}
We illustrate the  situation  by numerically solving \eqref{sang00} and \eqref{cexact} using the evaporation model of Eq.~\eqref{evaplaw}.
We take $D=3$, {$C_0=10$},  and $\epsilon=1$.  We use $\kappa=(5120 \pi/8) \times1/8$, where the last factor of $1/8$ converts from the usual three-dimensional expression for $ M=C/2$. The evaporation time in this case is $t_E=2.513 \times 10^5$. It is convenient to assume that the collapse started with the shell at rest (in this example at  {$R_0=10C_0$}), and simply vary the initial radial coordinate and the mass to cover all possible scenarios.

The rapid increase of $|\ddot R|$ after $\tau\sim\tau_1$ leads to  breakdown of the numerical integration of either the exact or approximate systems for sufficiently small $x$. Physically it means that after {the shell comes   to within a distance $\epsilon_*$ from} the Schwarzschild radius, Alice sees the  rest of the collapse and evaporation as nearly instantaneous: for the last point on the shell trajectory on Fig.~\ref{crash}, time dilation factor is $\dot T=3.05 \times 10^7$.

\begin{figure}[htbp]
\includegraphics[width=0.43\textwidth]{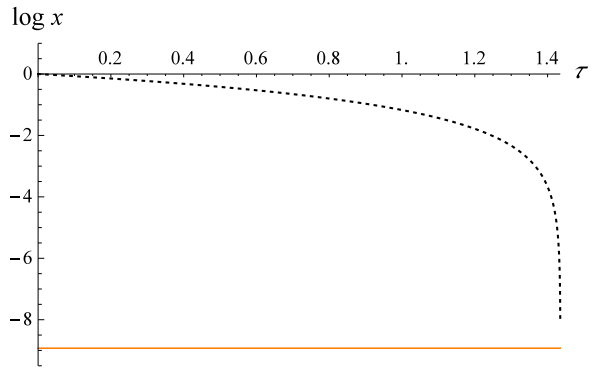}
\caption{Example: $D=3$, $C_0=10$, $R_0=100$. The exact solution (black dotted line) for the gap $x(\tau)=R(\tau)-r_\sg(\tau)\equiv R(\tau)-C(\tau)$ of the system of Eqs.~\eqref{sang00} and \eqref{cexact} is evaluated up to $\tau=1.434846531778$ ($t\approx89.5697$), where the numerical integration breaks down. The steady state value  $\epsilon_*=1/3\kappa C$ is shown as the orange line.   }
\label{crash}
\end{figure}

This problem is avoided if one rewrites the  equations in terms of Bob's time $t=T(\tau$), using
  \be
\dot R=R_T \dot T, \qquad \ddot R=R_{TT}\dot T^2+R_T \ddot T.
 \ee
Using Eq.~\eqref{met1} we obtain $\dot T$, and $\ddot T=(d\dot{T}/dT)\dot T$ (Appendix~\ref{rets}),
thus  allowing us to replace the derivative of $R$ over the proper time in Eq.~\eqref{sang00} by the derivatives over Bob's time.

The approximate equation for $x$ takes a particularly simple form
\be
\frac{dx}{dT}=\frac{dR}{dT}-\frac{dr_\sg}{dT}\approx-\frac{(D-2)x}{r_\sg}-\frac{dr_\sg}{dT}
\ee
and can be also obtained directly from Eq.~\eqref{xeps} by using the value of $|F|\approx|R_T|$ from Eq.~\eqref{fclose}. For $D=3$, $\epsilon_*=1/3\kappa C$, we find
\be
\frac{dx}{dt}=-\frac{x}{C}+\frac{1}{3\kappa C^2}. \label{xappro0}
\ee
When Eq.~\eqref{evaplaw} is substituted for $C(t)$  this equation has a solution in terms of the error function. The numerical solution to this equation that extends the exact solution to all times is depicted in  Fig.~\ref{eps-short}.

 \begin{figure}[htbp]
\includegraphics[width=0.5\textwidth]{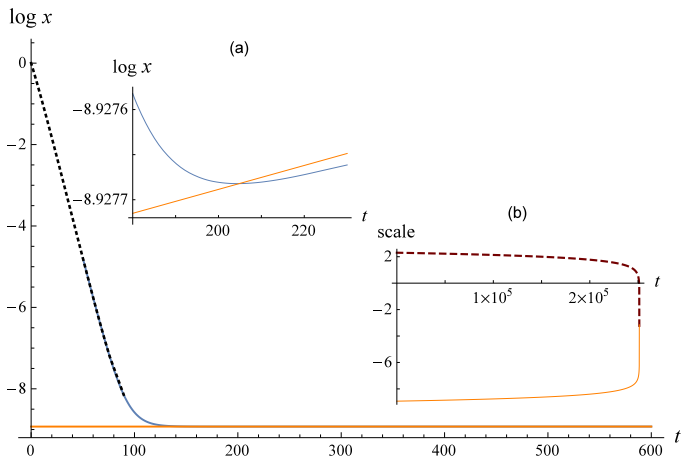}
\caption{Approach of  $x(t)$  to $\epsilon_*=1/3\kappa C$ (orange line) in the same setting. On this time scale $\epsilon_*\approx \mathrm{const}$. The plot is obtained as a combination of the solution  of the system of Eqs.~\eqref{sang00} and \eqref{cexact} (for $t<89.5697$, black dotted line) with the solution of the approximate  Eq.~\eqref{xappro0}   and with the adjusted initial conditions  ($t>50$, blue line). The inset (a) shows the approach of $x(t)$ to $\epsilon_*(t)$. After the moment $t_*$ when $x(t_*)=\epsilon_*(t_*)$ the distance from the Schwarzschild radius begins to increase, but it trails behind $\epsilon_*(t)$. The inset (b) shows the two scales of the problem: $\log C(t)$ (dashed dark red line) and $\log\epsilon_*(t)$ (orange line) through the evaporation. }
\label{eps-short}
\end{figure}

 \subsection{The core and the shell model}
  A more complicated scenario involves a massive spherically-symmetric core   {of radius $R_c$ with the Schwarzschild radius
 $r_\sg^\mathrm{c}=C_\mathrm{c}^{1/(D-2)}$.}  Assume a shell collapses from rest at some $R(0)$,
 such that  initially
 \be
 r_\sg^\mathrm{c}<R_\mathrm{c}<r_\sg<R
 \ee
where $C>C_\mathrm{c}$ yields the total gravitating mass of the system.

For a large core ($r_\sg^\mathrm{c}\ll R_\mathrm{c}$) an observer that is situated anywhere between it and the shell should not experience any deviation from   classical relativity. Hence the shell is the only source of Hawking radiation, and  Eqs.~\eqref{eq:eom-vaidya} and \eqref{sang00} are easily generalized (Appendix~\ref{coreshell}).   {Upon analysis we find} that the horizon still is not going to be reached, and the outcome of the evolution (the shell evaporates or crashes into the core) is determined by the two quantities: $x(\tau)$, and
\be
h(\tau):=r_\sg-R_\mathrm{c}.
\ee
We first observe that in any dimension the shrinking Schwarzschild radius will retreat inside the core, since at $\tau_\mathrm{r}$
\be
C(\tau_\mathrm{r})=R_\mathrm{c}^{D-2}>C_\mathrm{c},
\ee
indicating that the evaporation is not complete. Assuming that the shell is close to the Schwarzschild radius (and using the retarded Schwarzschild metric) we approximate $x(\tau)$ as $\epsilon_*(\tau)$ (Eq.~\eqref{epstars}), and the condition for crash becomes
\be
C^{1/(D-2)}(\tau_\mathrm{c})+\big(D(D-2)\kappa C(\tau_\mathrm{c})\big)^{-1}=R_\mathrm{c}.
\ee
The crash occurs if $C(\tau_\mathrm{c})>C_\mathrm{c}$. Otherwise the shell evaporates beforehand. For example, in $D=3$, the crash is possible if the core is large enough relative to its Schwarzschild radius,
\be
C(\tau_\mathrm{c})\approx R_\mathrm{c}-1/3\kappa R_\mathrm{c}>C_\mathrm{c}.
\ee
 This set-up provides a complimentary scenario to Ref.~\cite{ky:16}.

\section{Discussion}

Including the effects of pre-Hawking radiation dramatically modifies the evolution {of gravitationally collapsing matter}. In the examples we presented, for almost the entire evaporation time the shell stays very close to its Schwarzschild radius, but never crosses it. As a result, there are no trapped surfaces, no horizon and no singularity. The distance $\epsilon_*\propto C^{-1}$ grows as the shell evaporates {and
a comoving observer encounters an increased radiation flux with a natural cutoff --- a kind of tame firewall.
The distance from $r_\sg$ is within the trans-Planckian regime, but no more so than in the usual derivation of  Hawking radiation.  Both  horizon avoidance and its trans-Planckian scale support the idea that the paradoxical aspects of the black hole information problem originate in combining a sharp classical geometry with quantum fields \cite{brust:14, bht:17}.

It is quite reasonable that the mean field view of the collapse, radiation and backreaction breaks down well before the gap $\epsilon_*$ is reached. Quantum fluctuations become important and should be taken into account at least at the level of the stochastic gravity \cite{sto-g}. However, following this line of thought still implies that a classical picture of matter crossing a sharply defined surface is untenable, while quantum states that are associated with such a surface are not asymptotically reachable.

Despite ignoring the fluctuations our simple model is self-consistent.  The examples we have considered can be generalized to an arbitrary spherically-symmetric metric outside the shell \cite{us-new}.  Both avoidance of the horizon and a regularized firewall do not require any additional assumptions or exotic matter.

Currently popular  approaches to preserve unitary dynamics that are based on the analysis of matter alone  (such as  firewalls \cite{amps,bpz:13}, final projections \cite{lp:14}, or  ER=EPR \cite{ms:13,bv:14})  require both an horizon and a singularity and so are not applicable
if an event horizon does not form.  Our results also indicate the standard Penrose diagram that is used to illustrate the black hole creation and evaporation is inapplicable \cite{bht:17}.

Description of the entire spacetime in terms of a classical metric, even without the assumption that the Einstein
classical equations are be violated in the deep Planck region \cite{sthw:94}, results in disappearing of the event horizon, consitent with the expectations of \cite{ab:05}. While it is reasonable to assume that avoidance of a suitably defined Schwarzschild radius is a generic feature of a quantum collapse, lack of trapped surfaces and apparent horizon is most likely a consequence of the thin shell model. In general we expect appearance of the quasi-locally defined trapping horizons that should enable to contently discuss collapse, formation and evaporation of black holes \cite{ab:05}.

The smallness of $\epsilon_*$ seems to indicate that predictions of this model should be observationally indistinguishable from  pure classical collapse \cite{ast-bh}. However, it was recently shown that hypothetical  non-black-hole very compact objects  can have a very different quasinormal-mode spectrum  from that of  black holes, even in the limit of coinciding exterior metrics.   This difference is not manifested in the ringdown signal from a binary coalescence, but may become detectable  in   precision observations of the late-time ringdown signal \cite{pc:16}.

 Investigation of the energy-momentum tensor and its comparison with the results obtained from other considerations \cite{liberati, dfu:76, vsk:07,kmy:13,fw:99,ss:15,apt:16,ho:16}, as well as generalization to a non-zero pressure are natural extension of our results and will be published elsewhere.

Our model strengthens the point of view (see  \cite{h-shell,k:98,hr:15,ht:10})  that fully quantized joint gravity-matter dynamics must have unitary time evolution, particularly for systems that have a well-defined classical Hamiltonian,  and so there cannot be any overall information loss. Indeed, all current candidate theories of quantum gravity--- strings, loops, and foams --- are constructed as unitary theories.  However, an important  {unanswered question is how  entanglement (and more general types of quantum correlations) gets} distributed between the tripartite system of gravity--early modes--late modes. Similarly, if the fully formed horizon does not exist, it is important to  investigate how (if at all) the soft hair properties of black holes \cite{hps:16} are modified. Finally, by studying a realistic models it is important to understand if absence of the event horizon leads to astrophysically significant differences with the classical collapse.

\acknowledgments

  We thank Stefano Liberati and Bill Unruh for critical comments,  Paolo Pani, Tanmay Vachaspati and Yuki Yokokura  for useful suggestions. Robust discussions with Sabine Hossenfelder are gratefully acknowledged.  DRT and VB wish to thank Perimeter Institute and the Institute for Quantum Computing at the University of Waterloo for hospitality.  VB is supported by the Macquarie Research Fellowship scheme, RBM was supported by the Natural Sciences and Engineering Research Council of Canada.

\appendix
\section{ Angular coordinates in D+1 dimensions} \label{appa}
\label{angular-coordinates}

We choose the convention for labelling   angular coordinates such that the generalization of the $z$-axis is three spatial dimensions is referred as $x_1$, and there are $D-1$ angles $\phi_k$. We also introduce $\phi_D\equiv 0$ and set
$\prod_1^0 (*)\equiv 1$. Then
\be
x_k=r \cos\phi_k\prod_{l=1}^{k-1}\sin\phi_l.
\ee
 The spherical part of the metric  is
\be
r^2d\Omega_{D-1}:=r^2\left(d\phi_1^2+\sum_{a=2}^{D-1}\Big(\prod_{b=1}^{a-1}\sin^2\!\phi_b \Big)d\phi_a^2\right). \label{omd2}
\ee
The coordinates $y^a$ on the shell are $(\tau,\phi_k)$, and the induced metric is
\be
ds^2_\Sigma=-d\tau^2+R^2(\tau)d\Omega_{D-1}.
\ee

\section{ Thin shell collapse: outgoing Vaidya metric}  \label{curvature-v}

Dynamics of the shell is obtained via the second junction condition that equates the discontinuity of the extrinsic curvature with the surface energy-momentum tensor. Here we calculate the extrinsic curvature outside the shell. Expressions for the interior are analogous. Components of the extrinsic curvature are given by
\be
K_{\tau \tau} \coloneqq   -n_\alpha v^\alpha_{\,;\beta} v^{\beta} =  -n_\alpha a^\alpha, \qquad K_{\phi_k \phi_k} \coloneqq n_{\phi_k\,; \phi_k}.
\label{extr-curv}
\ee
 with $k=1,\ldots, D-1$ , where $v^\mu$ is the 4-velocity of $\Sigma$ and
  \be
 n_\mu=(-\dot R, \dot U,0,\ldots,0) \label{outnorm}
 \ee
is its outward-pointing unit normal.  The non-zero components of  the 4-acceleration  are
\be
a^0=\ddot{ U} - \frac{1}{2}F'\dot{ U}^2\, ,  \qquad a^1= \ddot R +\frac{1}{2} \left( F_U + FF'\right)\dot{ U}^2 + F' \dot{ U} \dot R.
\ee

Hence
\be
K^+_{\tau \tau} = \dot R \ddot{U} - \dot{U} \ddot R - \frac{3}{2} F' \dot{U}^2 \dot R  - \frac{1}{2} \left(F_U + F F'\right) \dot{ U}^3.
\ee
Calculating $\ddot{U}$ using Eq.~\eqref{udot}  we obtain
\begin{align}
\ddot{ U} =&\frac{\ddot R}{F}\left(\frac{\dot R}{\sqrt{F+ \dot R^2}}-1\right)+\frac{\dot R F'}{F}\left(\frac{1}{2\sqrt{F+ \dot R^2}}- \dot{ U}\right) \nonumber \\
&+ \frac{F_U \dot{ U}}{F} \left(-\dot{ U}+\frac{1}{2\sqrt{F+\dot R^2}}\right).
\end{align}
The extrinsic curvature can be decomposed into the part into a part that does not involve the derivative $F_U$ (the classical part), and the part that is proportional to it (the evaporating part), $K^+_{\tau \tau}=\mathcal{K}_c+\mathcal{K}_e$. We find
\be
\mathcal{K}_c=-\frac{2\ddot{R}+F'}{2\sqrt{\dot R^2+F}},
\ee
while
\begin{align}
{ \mathcal{K}_e} &= \dot{R}\frac{F_U \dot{ U}}{F} \left(\frac{1}{2\sqrt{F+\dot R^2}} -\dot{ U}\right)- \frac{1}{2} F_U \dot{ U}^3 \nonumber \\
& = -\frac{1}{2}\frac{F_U \dot{ U}^2}{\sqrt{F+\dot R^2}}.
\end{align}

The angular components --- $\phi_k \phi_k$ --- of the extrinsic curvature are calculated using the second relation in~\eqref{extr-curv}. Starting with $\phi_1 \phi_1$ --- $\theta \theta$ in 3+1-D --- we find
\begin{align}
K^+_{\phi_1 \phi_1}   =  \left(\dot R \, R  + F \dot{ U} \, R  \right) =   R \, \sqrt{F+\dot R^2},
\end{align}
where we have used the definition for $\dot{ U}$, and
\be
K^{\phi_1}_{+ \phi_1} = \frac{\sqrt{F+\dot R^2}}{R},
\ee
as expected. Going through a similar calculation we find that $K^{\phi_k}_{+ \phi_k}=  {\sqrt{F+\dot R^2}}/{R}$, for $k= 1, \dots, D-1 $.


Using the appropriate definitions of extrinsic curvature for the interior and the exterior regions, the components of the surface stress-energy tensor are
\begin{align}
S^\tau{}_\tau & =- \frac{1}{8\pi} \left[K^\tau_{+ \tau} - K^\tau_{- \tau} - \left([K^\tau{}_{\tau}] + \sum^{D-1}_{k=1} \left[K^{\phi_k}{}_{\phi_k}\right]\right) h^{\tau}{}_\tau\right] \nn \\
& = \frac{1}{8 \pi}(D-1) \left(\frac{\sqrt{F+\dot R^2}-\sqrt{1+\dot R^2}}{R}\right) = - \sigma, \label{tautau}
\end{align}
for the $\tau\tau$ component, and identical expressions such as
\begin{align}
S^{\phi_1}_{~\phi_1} & = \frac{1}{8 \pi}\left[\frac{ 2\ddot R+F'}{2\sqrt{F+\dot R^2}}-\frac{\ddot R}{\sqrt{1+\dot R^2}}  -\frac{1}{2}\frac{F_U \dot{ U}^2}{\sqrt{F+\dot R^2}}
\right.\nonumber\\
&\left. + (D-2) \left(\frac{\sqrt{F+\dot R^2}-\sqrt{1+\dot R^2}}{R}\right)\right] = 0,
\label{angular-rs}
\end{align}
 for all other components.

\subsection{Classical shell dynamics}
The classical equation \eqref{sangC} is obtained by suppressing the time-dependence of the metric and integrating Eq.~\eqref{tautau}. The shell surface density is conviniely expressed via the mass parameter $m$,
\be
\label{eq:C-non-evap}
C=2m\sqrt{1+\dot R^2} - \frac{m^2}{R^{D-2}}.
\ee
Fixing $C$ for the initial condition $\dot R(0)=0$ we get
\be
m=R_0^{D-2}\big(1-\sqrt{1-C/R_0^{D-2}}\big),
\ee
where $R(0)=R_0$. As a result, the equation of motion for $R$ is
\be
\frac{d R}{d \tau} = - \sqrt{\left(\frac{C}{2m}+\frac{m}{2 R^{D-2}}\right)^2-1}. \label{vel}
\ee

The initial position can be specified as $R_0=\big(\lambda C)^{1/(D-2)}$, for some $x>1$.Then
\be
m=\lambda C(1-\sqrt{1-1/\lambda}),
\ee
has the form that is independent of the dimension. In the limit $\lambda\rightarrow\infty$, $m\rightarrow C/2.$


\section{Inequivalence between the retarded Schwarzschild and the outgoing Vaidya metrics} \label{ineq}

Despite  similar reasoning behind the two metrics (and the intuitive feeling that $u$ and $\tit$ are directly related), there is no coordinate transformation between the coordinates $(t,r)$ and $(u,r)$ that will transform the metrics into each other if $C\neq\mathrm{const}$. Starting from the metrics \eqref{met0} and \eqref{met01} we have
\begin{align}
&g_{tt}=-f(u,r)\left(\frac{\pad u}{\pad t}\right)^2,\\
&g_{tr}=-f(u,r)\frac{\pad u}{\pad t}\frac{\pad u}{\pad r}-1 \frac{\pad u}{\pad t}\equiv 0,\\
&g_{rr}=-f(u,r)\left(\frac{\pad u}{\pad r}\right)^2-2\frac{\pad u}{\pad r}.
\end{align}
Hence we find
\be
\frac{\pad u}{\pad r}=-\frac{1}{f(u,r)},
\ee
and consequently $g_{rr}=1/f(u,r)$. To enforce $g_{tt}=-f(u,r)$,
\be
\frac{\pad u}{\pad t}=\pm 1,
\ee
should hold. However,  the integrability condition implies
\be
\frac{\pad^2 u}{\pad t\pad r}=\frac{1}{f^2(u,r)}\frac{\pad f}{\pad t}=\frac{\pad^2 u}{\pad r \pad t}=0,
\ee
that holds only in a stationary spacetime.

\section{Thin-shell collapse: the retarded Schwarzschild metric}\label{rets}

The derivation proceeds analogously to case of the outgoing Vaidya metric. The outward-pointing normal is
 \be
 n_\mu=(-\dot R, \dot T,0,\ldots,0),
 \ee
and the non-zero components of  the 4-acceleration components are
\begin{align}
a^{t} = \ddot{T}(\tau) + \frac{1}{2} F^{-1} \left[ \left(F^{-1}\right)_t \dot{R}^2 + 2 \dot{R} \, \dot{T}   F' + F_t \dot{T}^2\right], \\
a^{r} = \ddot{R}+ \frac{1}{2} F \left[ \left(F^{-1}\right)'\dot{R}^2 + 2 \dot{R} \, \dot{T}  \left(F^{-1}\right)_t + F' \dot{T}^2\right].
\end{align}
Hence
\begin{align}
K^+_{\tau \tau} &=  \dot{R}\, \ddot{T} - \dot{T} \ddot{R} +\frac{3}{2} F^{-1} F' \dot{T} \dot{R}^2 - \frac{1}{2} FF' \dot{T}^3 \nonumber \\
 &+  \frac{1}{2} F^{-1} F_t \dot{R} \, \dot{T}^2 + \frac{1}{2} F^{-1} (F^{-1})_t\dot{R}^3 - F (F^{-1})_t \dot{R} \dot{T}^2.
\end{align}
Calculating $\ddot{T}$ using Eq.~\eqref{tdot}  we obtain
\be
\ddot{T} =  \frac{2 \dot{R} \ddot{R} F - F F' \dot{R} - 2 F' \dot{R}^3 }{2 F^2 \sqrt{F +\dot{R}^2}}-\frac{F_t}{2F^3}(F+2\dot R^2).
\ee
Substituting $\ddot T$ and with further simplifications we obtain
\be
K^+_{\tau \tau} = -\frac{2 \ddot R + F'}{2\sqrt{F + \dot R^2}} + \frac{F_t \dot R}{F^2}.
\ee
The angular components of the extrinsic curvature are
\be
\label{eq:K-angle}
K_{\phi_k\phi_k}^+=n_{\phi_k;\phi_k}^+=\half g^{rr}g_{\phi_k\phi_k,r}n_r=F(R,T)R\dot{T}\prod_{b=1}^{k-1}\sin^2\!\phi_b.
\ee
(note that $e^\mu_a=\delta^\mu_{a+1}$ for $\mu=2,\ldots D$). From the above equations one can therefore see that $K_{\phi_k\phi_k}^+$ is not affected by the evaporation.

The derivatives of $\tit$ are found via
\begin{align}
\frac{\pad}{\pad t}(t-\tit)&=1-\frac{\pad\tit}{\pad t}=\frac{\pad}{\pad t}\int_{R(\tit)}^r\!\frac{dr'}{f(\tit,r)}\nonumber \\
&=-\frac{1}{F\big(\tit,R(\tit)\big)}\frac{d R}{d\tit}\frac{\pad\tit}{\pad t}
-\int_{R(\tit)}^r\!\frac{dr'}{f^2(\tit,r)}\frac{\pad f}{\pad\tit}\frac{\pad \tit}{\pad t}.
\end{align}
In the limit $r\rightarrow R(\tit)$ (and, i.e., $T=t\rightarrow\tit$) we have
\be
1-\frac{\pad\tit}{\pad t}=-\frac{1}{F\big(\tit,R(\tit)\big)}\frac{d R}{d\tit}\frac{\pad\tit}{\pad t},
\ee
hence
\be
\frac{\pad\tit}{\pad t}\Big|_\Sigma=\frac{1}{1-R_T/F\big(T,R(T)\big)}.
\ee
Similarly, in the same limit we obtain
\be
-\frac{\pad\tit}{\pad r}\Big|_\Sigma=\frac{1}{F\big(T,R(T)\big)}-\frac{1}{F\big(\tit,R(\tit)\big)}\frac{d R}{d\tit}\frac{\pad\tit}{\pad r}\Big|_{r=R(t)}.
\ee
Hence
\be
\frac{\pad\tit}{\pad r}\Big|_\Sigma=-\frac{1}{F\big(T,R(T)\big)-R_t}.
\ee

 Changing to the coordinate time parametrization of the equation of motion requires the expressions for $\dot{T}$ and $\ddot T$. From Eq.~\eqref{met0} we see that
\be
-d\tau^2=-FdT^2+R_T^2 dT^2/F,
\ee
 so
\be
\dot{T}=\sqrt{\frac{F}{F^2-R_T^2}},
\ee
and
\be
\ddot T=\frac{d\dot T}{dT}\dot T =\frac{2F R_T R_{TT}-\big(F^2+R_T\big)^2dF/dT}{2\sqrt{F\big(F^2-R_T^2}\big)}\dot T.
\ee

\section{ Page's formula in $D$ spatial dimensions} \label{page-d}

The flux of radiation in $D$ spatial dimensions is
\be
J=\sigma_D T^{D+1}, \label{jtd}
\ee
where $\sigma_D$ is the $D$-dimensional Stefan-Boltzmann constant. For comparison:
\be
\sigma_3=\frac{\pi^2}{60\hbar^3 c^2}, \qquad  \sigma_D=\frac{g_D S_{D-2} D!\zeta(D)}{2(2\pi)^D }\frac{1}{\hbar^D c^{D-1}}.
\ee
The total radiated power is
\be
L=\gamma \sigma_D T_H^{D+1} A_D,
\ee
where $\gamma$ is the number of species,  and $A_D=S_{D-1}r_g^{D-1}$, with the Schwarzschild radius
\be
r_g= C_D^{\frac{1}{D-2}}.
\ee
Note that in 3D the constant $\sigma_3$ is already calibrated to include two polarizations of photons.

The mass is given by
\begin{equation}
M=\frac{C_D (D-1) S_{D-1}}{16 \pi {G_{(D+1)}}}=:\alpha_D C_D. \label{cmass}
\end{equation}
The horizon area is
\be
A_D=S_{D-1}C_D^{\frac{D-1}{D-2}}=S_{D-1}\left(\frac{16 \pi G_{D+1}M}{(D-1)S_{D-1}}\right)^{\frac{D-1}{D-2}}.
\ee
The Hawking temperature  dimensions is defined as
\begin{equation}
T=\frac{f'(r_g)}{4\pi},
\end{equation}
which in $D+1$ spacetime dimensions is in
\be
T_D=\frac{1}{4 \pi}\frac{D-2}{\sqrt[{D-2}]{C_D}}.
\end{equation}

The rate of change of $C_D$ according to an observer at the spatial infinity is
\begin{align}
\frac{d C_D}{d t} &= -   \left(\frac{16 \pi  {G_{(D+1)}}}{(D-1) S_{D-1}}\right)^{\frac{D}{D-2}} P(D)\frac{1}{C^{\frac{2}{D-2}}} \\&=\alpha_D^{{-D/(D-2)}}P(D)C_D^{-2/(D-2)}=:-\varpi(D)C_D^{-2/(D-2)}.\nonumber
\end{align}
We will write the evaporation time
\be
t_E=\kappa C_0^{\frac{D}{D-2}} \qquad \kappa=\frac{1}{\varpi(D)}\frac{D-2}{D},
\ee
where    $C_0$ is the initial value of $C_D$.

\section{ A thin shell collapsing on a massive core}\label{coreshell}

In the case of a thin shell collapsing on a core, the metric of the inner region is Schwarzschild as well, with $f_-= 1- {C_c}/{r^{D-2}}$.
The equations of motion combine features of the proceeding cases, and are calculated in terms of the surface stress-energy tensor $S^a{}_b$. The time-time component condition now reads
\be
S^\tau{}_\tau= \frac{1}{8 \pi } (D-1)  \left(\frac{\sqrt{F+\dot R ^2} - \sqrt{F_-+ \dot R^2}}{R} \right) =  -\sigma,
\ee
while the angular component equations become
\begin{widetext}
\be
\frac{1}{8 \pi }\left[\frac{2\ddot R + F'}{2\sqrt{F+\dot R^2}} - \frac{2\ddot R+F'_-}{2\sqrt{F_-+\dot R^2}} - \frac{F_t \dot R}{F^2}+ (D-2)\frac{\sqrt{F + \dot R^2} - \sqrt{F_-+\dot R^2}}{R}  \right] = 0. \label{sang}
\ee
\end{widetext}
Evaporation affects only the metric outside, i.e.
\be
F'=\frac{C (D-2)}{R^{D-1}}-\frac{d C}{d \tit}\frac{\pad\tit}{\pad t}\frac{1}{R^{D-2}},
 \qquad
F'_-=\frac{C_\mathrm{c} (D-2)}{R^{D-1}},
\ee
where according to Eq.~\eqref{cmass}
\be
C=M/\alpha_D, \qquad C_\mathrm{c}=\Mc/\alpha_D.
\ee
On the other hand, only the outside term
\be
F_t= -\frac{d C}{d\tit}\frac{\pad\tit}{\pad t}\frac{1}{R^{D-2}},
\ee
enters the equation.  The qualitative behaviour of the collapsing shell is the same as that for a collapsing shell with no core. The only effect of the core on the dynamics of the system is to accelerate the rate of evaporation.

\subsection{Asymptotic dynamics analysis}
We again consider $x= R- r_\sg$. By expanding $\ddot R$ in inverse powers of $C$ and $x$, and only taking the leading terms that diverge when $x\rightarrow 0$
\be
\ddot{R} \approx \frac{\dot R^2 \sqrt{\dot R^2 + 1}}{2 D (D-2)\kappa x ^2 \, C \left(\dot R + \sqrt{1+\dot R^2}\right)} +\mathcal{O}\left(x^{-1}\right),
\ee
the dominant terms of the acceleration are the same as for Eq.~\eqref{evaptau}. We can still use the approximations
\be
F\approx \frac{(D-2)x}{C^{1/(D-2})}\equiv \frac{(D-2)x}{r_\sg}, \qquad \dot T \approx -\frac{r_\sg}{(D-2)x} \dot R,
\ee
and once$ |\dot R| \gtrsim 2$
\be
\label{rdotdot}
\ddot R \approx -\frac{\dot R^4}{D(D-2 \kappa C x^2)}.
\ee
Similar to the case of a collapsing thin shell --- without the core --- we have that
\be
\dot R = \frac{d R}{dt} \dot T \approx -\frac{d R}{dt } \frac{\dot R}{F}.
\ee
Since in this regime
\be
R_{\tilde t}|_{\Sigma}= R_t\equiv R_T\approx - F, \qquad \frac{\partial \tilde t}{\partial t} \bigg|_{\Sigma} \approx \frac{1}{2}, \qquad \frac{\partial \tilde t}{\partial r} \bigg|_{\Sigma} = -\frac{1}{2F}.
\ee
The evaporation equation in $D+1$-dim (as seen from Alice)
\be
\frac{d C}{d \tau} = \frac{d C}{d \tilde t} \left( \frac{\partial \tilde t}{\partial t} \dot T + \frac{\partial \tilde t}{\partial r} \dot R \right),
\ee
can be approximated as
\be
\label{cdot}
\dot C \approx -\frac{dC}{d \tilde t} \frac{\dot R}{F} = \frac{1}{D \kappa} \frac{\dot R}{C^{1/(D-2)}}.
\ee
Eqs~\eqref{rdotdot} and~\eqref{cdot} give us the dynamics of the shell and of the Schwarzschild radius. Hence
\be
\dot r_\sg=\frac{1}{D-2}\frac{\dot C}{C^{(D-3)/(D-2)}}=\frac{1}{D(D-2)\kappa } \frac{\dot R}{C x}.
\ee
that means the gap $x$ evolves according to
\be
\dot x =  \dot R \left(1 - \frac{1}{D(D-2)\kappa} \frac{1}{C x}\right) = |\dot R| \left(\frac{1}{D(D-2)\kappa} \frac{1}{C x} - 1\right),
\ee
when $x< \epsilon_*$, with
\be
\epsilon_* (\tau) = \frac{1}{D(D-2)\kappa C(\tau)}.
\ee
we are guaranteed $\dot x > 0$, i.e. the shell does not cross the Schwarzschild radius.


\begin{thebibliography}{99}
\bibitem{ll2} L. D. Landau and E. M. Lifshitz, \textit{Classical Theory of
Fields} (Reed International, Oxford, 1975).


\bibitem{kiefer:07} C. Kiefer, \textit{Quantum Gravity}, (Oxford University Press, Oxford, 2007).
\bibitem{bd:82} N. D. Birrel and P. C. W. Davies, \textit{Quantum Fields in Curved Space} (Cambridge University Press, Cambridge, 1986).
\bibitem{modern}  R. B. Mann, \textit{Black Holes: Thermodynamics, Information, and Firewalls} (Springer, New York, 2015).
\bibitem{modern2}  D. Harlow,  {Rev. Mod Phys.} \textbf{88}, 015002 (2016).
\bibitem{h:74} S. W.  Hawking,  {Nature}  \textbf{248}, 30 (1974).
\bibitem{heffler:03} A. D. Heffler, Rep. Prog. Phys. \textbf{66}, 943 (2003).
\bibitem{page:76} D. N. Page,  Phys. Rev. D \textbf{13}, 198 (1976).
\bibitem{ab:05}  A. Ashtekar and M. Bojowald, Class. Quant. Grav \textbf{22}, 3349 (2005).

\bibitem{wald:01lrr}  R. M. Wald,  {Living. Rev. Rel.} \textbf{4}, 6 (2001).
\bibitem{mathur:09} S. D. Mathur, Class. Quantum. Grav. \textbf{26}, 224001 (2009).
\bibitem{new} W. G. Unruh and R. M. Wald, arXiv:1703.02140 (2017);  D. Marolf, arXiv:1703.02143 (2017).
\bibitem{visser:08} M. Visser, PoS BHs,GRandStrings 2008:001 (2008),   	arXiv:0901.4365v3.

\bibitem{ger:76} U. H. Gerlach, Phys. Rev. D \textbf{14}, 1479 (1976).
\bibitem{haj:86} P. H\'{a}j\'{\i}\v{c}ek,  Phys. Lett. B \textbf{182}, 309 (1986); P. H\'{a}j\'{\i}\v{c}ek, Phys. Rev. D \textbf{36}  1065 (1987).
\bibitem{acvv:01} G. L. Alberghi,  R. Casadio, G. P. Vacca,  and G. Venturi, Phys. Rev. D  \textbf{64}, 104012 (2001).
\bibitem{liberati} C. Barcel\'{o}, S. Liberati, S.  Sonego, M. Visser, {Class. Quant. Grav.}   \textbf{23}, 5341 (2006);
  C. Barcel\'{o}, S. Liberati, S.  Sonego, M. Visser, {Phys. Rev. D} \textbf{77}, 044032 (2008).
\bibitem{vsk:07} T. Vachaspati, D. Stojkovic, and  L. M. Kraus, {Phys. Rev. D} \textbf{76}, 024005 (2007); T. Vachaspati and D. Stojkovic, Phys. Lett. B \textbf{663}, 107 (2008).
\bibitem{kmy:13} H. Kawai, Y. Matsuo, and Y. Yokokura,   {Int. J. Mod. Phys. A} \textbf{28}, 1350050 (2013); H. Kawai and Y. Yokokura,
Int. J. Mod. Phys. A \textbf{30}, 1550091 (2015).
\bibitem{ho:16a} P.-M. Ho, {Nucl. Phys. B} \textbf{909},   394 (2016).


\bibitem{h:76} S. W. Hawking,   {Phys. Rev. D} \textbf{14}, 2460 (1976).

\bibitem{thes} D. R. Terno, \textit{Quantum Information and Relativity Theory}, Ph.D. thesis, (Technion, Haifa, 2003).
 \bibitem{pt:04} A. Peres and D. R. Terno, Rev. Mod. Phys. \textbf{76}, 93 (2004).
 \bibitem{d:10} A. Dragan,  arXiv:1610.07839 (2010).

 \bibitem{sto-g} J. F. Donoghue, Phys. Rev. D \textbf{50}, 3874 (1994); C.P.
Burgess, Living Rev. Relativity \textbf{7}, 5 (2004).
\bibitem{eft-g} B.-L. Hu and E. Verdauger, Class. Quant. Grqav \textbf{20}, R1 (2003); B.-L. Hu and E. Verdauger, Liv. Rev. Relativity \textbf{11}, 3 (2008).
\bibitem{amps} A.  Almhieri, D. Marolf,   J. Polchinski, and J. Sully, {JHEP} \textbf{02}, 062 (2013).
\bibitem{hr:15} H. M. Haggard and C. Rovelli, Phys. Rev. D. \textbf{92}, 104020 (2015).
\bibitem{bht:17} V. Baccetti,  V. Hussain, and D. R. Terno, Entropy \textbf{19}, 17 (2017).
\bibitem{us-new} V. Baccetti, R. B. Mann, and D. R. Terno, arXiv:1703.09369 (2017).
\bibitem{poisson} E. Poisson, \textit{A Relativist's Toolkit}, (Cambridge University Press, Cambridge, 2004).
\bibitem{hor:12} G. T. Horowitz, ed., \textit{Black Holes in Higher Dimensions}, (Cambridge University Press, Cambridge, 2012).
\bibitem{isr:66} W. Israel, Nuovo Cimento \textbf{44B}, 1 (1966); \textbf{48B}, 463 (1967).


\bibitem{dfu:76} P. C. W. Davies, S. A. Fulling, and W. G. Unruh, Phys. Rev. D \textbf{13}, 2720 (1976).

\bibitem{h-shell} P. H\'{a}j\'{\i}\v{c}ek, Nucl. Phys. B \textbf{603}, 555 (2001); M. Ambrus and P. H\'{a}j\'{\i}\v{c}ek, Phys. Rev. D \textbf{72}, 064025 (2005).
    \bibitem{ss:15} A. Saini and D. Stojkovic, Phys. Rev. Lett. \textbf{114}, 111301 (2015).

\bibitem{vai:51} P. C. Vaidya, Phys. Rev. \textbf{83}, 10 (1951).
\bibitem{fw:99} M. K. Parikh and F. Wilczek, Phys. Lett. B \textbf{449}, 24 (1999).
\bibitem{ft:08} F. Fayos and R. Torres, Class. Quant. Grav. \textbf{25}, 175009 (2008).
\bibitem{pz:84} D. Page and W. H. Zurek, Phys. Rev. D \textbf{29}, 628 (1984).
\bibitem{z-ap} H. D. Zeh, \textit{The Nature and Origins of Time-Asymmetric Spacetime Structures}, in A. Ashtekar and V. Petkov (eds.), \textit {Springer Handbook of Spacetime} (Springer, Berlin, 2014), p. 185.

\bibitem{bbg:11} L. C. Barbado, C. Barcel\'{o}, and L. J. Garay, Class. Quant. Grav. \textbf{28}, 125021 (2011).
\bibitem{ky:16} H. Kawai and Y. Yokokura, Phys. Rev. D. \textbf{93}, 044011 (2016).

\bibitem{brust:14}  R. Brustein,   Fortschr. Phys. \textbf{62}, 255 (2014).

\bibitem{ast-bh} R. Narayan and J. E. McClintock, New Astr. Rev. \textbf{51}, 733 (2008); M. Abramowicz and C. Fragile, Living Rev. Relativity \textbf{16}, 1
(2013).
\bibitem{pc:16} V. Cardoso, E. Franzin, and P. Pani, Phys. Rev. Let. \textbf{116}, 171101 (2016); V. Cardoso, S. Hopper, C. F. B. Macedo, C. Palenzuela, and  P. Pani, Phys. Rev. D \textbf{94}, 084031 (2016).

\bibitem{sthw:94} C. R. Stephens, G. 't Hooft, and B. F. Whiting,   Class. Quant. Grav. \textbf{11}, 621 (1994).

\bibitem{bpz:13} S. L. Braunstein, S. Pirandola, and K.  \.{Z}yczkowski, Phys. Rev. Lett. \textbf{110}, 101301 (2014).
\bibitem{lp:14} S.  Lloyd, and J.  Preskill, {JHEP} \textbf{1408}, 126 (2014).
\bibitem{ms:13}  J. Maldacena,  and L.  Susskind, {Fortschr. Phys.} \textbf{61}, 781 (2013).
\bibitem{bv:14}  J. C. Baez and J. Vicary,  {Class. Quant. Grav.} \textbf{31}, 214007 (2014).


\bibitem{apt:16} S. Abdolrahimi,  D. N. Page, and C. Tzounis, arXiv:1607.05280 (2016).
\bibitem{ho:16} P.-M. Ho, Class. Quant. Grav.  \textbf{34}, 085006 (2017).
\bibitem{k:98} B. S. Kay,  Class. Quant. Grav. \textbf{15}, L89 (1998);  arXiv:hep-th/9802172 (1998).
\bibitem{ht:10} V. Husain and D. R. Terno, Phys. Rev. D \textbf{81},  044039 (2010).
\bibitem{hps:16} S. W. Hawking, M. J. Perry, and A. Strominger, Phys. Rev. Lett. \textbf{116}, 231301 (2016).











\end{thebibliography}
\end{document}